\documentclass[10pt,aps,twocolumn,showpacs]{revtex4}
\usepackage{epsfig}
\usepackage{bbm}
\usepackage{amssymb}
\usepackage{amsmath}

\newcommand{\lbfig}[1]{\refstepcounter{fig} \label{#1} }
\newcounter{fig}
\newcommand{\nc}{\newcommand}
\nc{\be}{\begin{equation}}
\nc{\ee}{\end{equation}}
\nc{\bea}{\begin{eqnarray}}
\nc{\eea}{\end{eqnarray}}
\nc{\nn}{\nonumber}

\begin{document}
\leftline{\hspace{4.5in} CERN-TH/2003-194, HD-THEP-03-37}

\vskip 0.2in
\title{Axial Currents from CKM Matrix CP Violation and Electroweak
        Baryogenesis}

\author{Thomas Konstandin$^*$, Tomislav Prokopec$^*$
                       and 
                   Michael G. Schmidt
       }
\email[]{T.Konstandin@thphys.uni-heidelberg.de}
\email[]{T.Prokopec@thphys.uni-heidelberg.de}
\email[]{M.G.Schmidt@thphys.uni-heidelberg.de}

\affiliation{Institut f\"ur Theoretische Physik, Heidelberg University,
             Philosophenweg 16, D-69120 Heidelberg, Germany}

\date{\today}

\begin{abstract}
  The first principle derivation of kinetic transport equations suggests that 
a CP-violating mass term during the electroweak
phase transition can induce axial vector
currents. Since the important terms are of first order in gradients
there is a possibility to construct new rephasing invariants that 
are proportional to the CP phase in the
Cabibbo-Kobayashi-Maskawa matrix and to circumvent the 
upper bound of CP-violating contributions in the Standard Model, the 
Jarlskog invariant. Qualitative arguments are given that these new 
contributions still fail to explain electroweak baryogenesis in extensions of 
the Standard Model with a strong first order phase transition. 
\end{abstract}

\pacs{
98.80.Cq,  
11.30.Er,  
11.30.Fs   
} 

\maketitle

%
%

\section{Introduction}

Following the seminal work~\cite{Kuzmin:1985mm}  about electroweak baryogenesis (EWB) 
many models have been proposed in the last years, that intend to
explain the baryon asymmetry of the universe (BAU) by sphaleron processes that
couple to an axial quark current during a first order electroweak phase transition (EWPT).
The main reason that this topic attracted such an
attention is,
that the related elementary particle physics is accessible to experiments these days.

However all models depend on extensions of the Standard Model (SM)
since the SM fails on the following grounds:

\begin{itemize}

\item Lack of CP violation

Since the only source of CP violation in the Standard Model is the
Cabibbo-Kobayashi-Maskawa (CKM) matrix (apart from the neutrino mass matrix,
which provides an even tinier source of CP violation) one has to face that it is 
too weak to account for the observed magnitude of BAU.

\item First order phase transition

Sakharov~\cite{Sakharov:1967} pointed out that baryogenesis necessarily
requires non-equilibrium physics. The expansion of the universe is too slow
at the electroweak scale and one needs bubble nucleation during a first
order EWPT. The phase diagram of the Standard Model is studied
in detail~\cite{Kajantie:1996qd,Kajantie:1996mn},
and it is well known that there is no first order phase transition
in the Standard Model for the experimentally allowed Higgs mass. 

\item Sphaleron bound

To avoid washout after the phase transition, the {\it vev} of the broken
Higgs field has to meet the criterion $ \langle \Phi \rangle \gtrsim T_c$,
i.e. a strong first order phase transition. This results
in the Shaposhnikov bound on the Higgs
mass~\cite{Shaposhnikov:1986,Shaposhnikov:1987}.

\end{itemize}

In the following we will address the first point - the lack of sufficient 
CP violation. The strong first order phase transition is assumed to 
occur at about $T_c\simeq 100$ GeV and is parametrised by the velocity
of the phase boundary (wall velocity) $v_w$ and its thickness $l_w$. 
It may be induced by adding massive scalars and gauge fields to the SM.

A first attempt to account for the BAU within the SM was given by Farrar and
Shaposhnikov~\cite{Farrar:sp}.
Their method was based on reflection coefficients in the thin wall regime and 
the need of different diagonalization matrices in the broken and the unbroken phases.
However, it has been argued, that the fermion damping by gluons annihilates the
coherent modes too fast to account for an axial quark current in the 
wall~\cite{Gavela:ds,Huet:1994jb}.

On the other hand one important effect has been neglected
by the assumption that the wall is infinitely thin.
The use of a continuous wall profile in the WKB 
approach~\cite{JoyceProkopecTurok:1994}
leads to a dependence of the 
dispersion relation on the CP-violating phase of complex mass terms.
This effect is in strong contrast to the coherent generation
of axial fermion currents
as discussed in~\cite{Farrar:sp},  
since in the former case the wall produces a 
CP sensitive mass of the fermions and the damping is {\it required} to
convert this slight mass change into a displacement in the fermionic 
distribution functions. 

A first principle derivation of this dispersion relation and the 
associated transport equation in the Schwinger-Keldysh formalism 
was given in~\cite{Kainulainen:2001cn}. 
This formalism is shortly reviewed in the next two sections.
Content of the present publication is to generalise this method to
several flavors and to the Standard Model type case, in which
the Cabibbo-Kobayashi-Maskawa matrix is the source of 
CP violation. A similar mechanism, coherent baryogenesis, where mixing
of fermionic flavours was used to generate baryon asymmetry, has been
recently proposed~\cite{Garbrecht:2003mn}.
\section{The Kadanoff-Baym Equations} 

We start our analysis with the exact Schwinger-Dyson equation
for two point functions in the closed time path (CTP) formalism.
After a transformation into 
the Wigner-space they read (our notation
is the usual one~\cite{Kainulainen:2002sw}; 
flavor and spinor indices are suppressed)
\begin{eqnarray}
&& \hskip -0.50 cm {\rm e}^{-i\Diamond} \{ S_0^{-1} - \Sigma_R , S^< \} 
-  {\rm e}^{-i\Diamond} \{ \Sigma^< , S_R \} \nonumber\\
&&\hskip 1.50 cm =\; \frac{1}{ 2} {\rm e}^{-i\Diamond} \{ \Sigma^< , S^> \} 
- \frac{1}{ 2} {\rm e}^{-i\Diamond} \{ \Sigma^> , S^< \}
\quad
\nonumber \\ 
&& \hskip -0.50 cm {\rm e}^{-i\Diamond} \{ S_0^{-1} - \Sigma_R , {\cal A} \} 
-  {\rm e}^{-i\Diamond} \{ \Sigma_A , S_R \}
 = 0
\,,
\label{KB}
\end{eqnarray}
where $S_0^{-1} 
$
is the inverse free propagator,
and we have used the definitions and relations
\begin{eqnarray}
 && \hskip -2cm S^{\bar t}:=S^{--}, \;\; 
                S^t:=S^{++}, \;\; 
                S^<:=S^{+-}, \;\; 
                S^>:=S^{-+}, 
\nonumber \\
 {\cal A}&:=&\frac{i}{2}(S^>-S^<),   \nonumber \\
 S_R&:=&S^t-\frac{1}{2}(S^<+S^>),  \nonumber \\
 \Sigma_A&:=& \Gamma = \frac{i}{2}(\Sigma^>-\Sigma^<),  \nonumber \\
\Diamond\{A,B\} &:=& \frac{1}{2} \left( \partial_{X} A \cdot \partial_{k} B
-\partial_{k} A \cdot \partial_{X} B \right),
\end{eqnarray}
and $--,-+,+-,++$ denote the four propagators of the Keldysh $2\times 2$ 
matrices and all functions depend on the momentum $k_\mu$ and 
the average coordinate $X_\mu$.

Up to this point these equations are formally exact and hard to solve. They simplify when expanded 
in gradients. The terms on the left hand side will be expanded up to first order, whereas the 
collisional sources on the right hand side vanish in equilibrium and are 
just taken up to zeroth order. The expansion parameter is formally
$\partial_X/k$, which close to equilibrium and for typical thermal 
excitations reduces to $(l_w \,T)^{-1}$. Here $T$ denotes the temperature
and $l_w$ is the phase boundary thickness of the bubbles of 
the first order electroweak phase transition.

We will perform the calculation in the quasiparticle limit. 
In practice this means, that we 
neglect the last terms on the left hand side, which contain $S_R$ and 
give rise to a Breit-Wigner type spectral function~\cite{Henning:sm}.

The equations up to first order in gradients are therefore:
\begin{eqnarray}
 && \hskip -0.5cm (S_0^{-1} - \Sigma_R) S^< 
 - \frac {i}{2} \partial_X (S_0^{-1} - \Sigma_R)\cdot \partial_k S^< 
\label{sd1} \\
&& \hskip 0.5cm
  +\; \frac{i}{ 2} \partial_k (S_0^{-1} - \Sigma_R)\cdot \partial_X S^<
  = \frac{1}{ 2}  \Sigma^<  S^>  - \frac{1}{ 2}  \Sigma^>  S^<   
\quad\;
\nonumber \\
 && \hskip -0.5cm (S_0^{-1} - \Sigma_R) {\cal A} 
 - \frac {i}{2} \partial_X (S_0^{-1} 
 - \Sigma_R)\cdot \partial_k {\cal A}
\nonumber \\   
&& \hskip 0.5cm
 +\; \frac{i}{ 2} \partial_k (S_0^{-1} - \Sigma_R)\cdot \partial_X {\cal A}  
\, =\,  0
\,. 
\label{sd2}
\end{eqnarray}
As additional simplification, one can treat all appearing self-energies 
as being in equilibrium --
which is more crude than a strict linear response approximation,
valid close to thermal equilibrium and whose implementation 
would imply additional integral terms -- such that
by using the Kubo-Martin-Schwinger condition and the thermal fermionic 
distribution function 
$f_F(k \cdot u) = [\exp{(k \cdot u / T)}+1]^{-1} $ 
($u^\mu=\gamma_w(1,0,0,v_w)$ denotes the plasma vector in the wall frame)
the collision terms in the first equation can be transformed into

$$ \frac{1}{ 2} \Sigma^< S^>  - \frac{1}{ 2} \Sigma^> S^<  = 
 \frac{1}{ 2} \Sigma_A (S^< - f_F(k \cdot u) \, \cal{A}). $$

We will not solve the full transport equations, but only look for the
appearing CP-violating source terms. The explicit form of the collision term
will not be discussed and is generally denoted by $Coll.$, even though
it can contain CP-violating contributions~\cite{Kainulainen:2002sw}.
We expect to obtain the usual classical transport equation to order
$\hbar^0$ and CP-violating effects to order $\hbar^1$, which
appear at the first order in gradients. 

\section{Model with CP-Violating Complex Mass}

To start with~\cite{Kainulainen:2001cn} we add a pseudoscalar imaginary
mass term to the normal Dirac operator. 
The inverse propagator in a convenient coordinate system, in the wall frame
in which a particle moves perpendicular to the wall $(\vec k_\|=0)$ and 
in the case of a stationary wall, reads 
$$ 
(S_0^{-1} - \Sigma_R) \rightarrow  \tilde k^0 \gamma_0 + k^3 \gamma_3 
 + m_R(X_3) + i m_I(X_3) \gamma_5,
$$
where $\tilde k^0 := {\rm sign}[k^0]({k^0}^2 - {\vec k_\|}^2)^{1/2}$.
The wall velocity will enter 
in the boundary conditions of $S^<$.

Although $m_R$ and $m_I$ are Lorentz scalars, they can appear in our solutions only in certain 
combinations. Since a chiral transformation will change the complex mass by a constant phase,
but should not have any physical relavance, only the following 
terms are possible up to second order in gradients 
(prime means differentiation with respect to $X_3$,
$m_R + i \, m_I = m \, {\rm e}^{i\theta}, m^2 = m_R^2 + m_I^2$)
\bea
&(m^2)^\prime, \theta^\prime m^2, (m^2)^{\prime\prime}, \theta^{\prime\prime} m^2, 
(m^2)^\prime \, \theta^\prime m^2. &\nonumber
\eea
The first CP-violating effect will therefore be at least of first order in gradients.

Since the inverse propagator commutes with the spin projector
$P_s = \frac{1}{2}(1 + s \gamma_0 \gamma_3 \gamma_5 )$, spin is conserved and 
the spin diagonal entries can be written in the block-diagonal form in 
spin,
\bea
&S^{<s,s} =P_s S^< P_s = 
P_s ( s^s_0 \gamma_0 + s^s_1 1 + s^s_2 i\gamma_5 + s^s_3 \gamma_3 ),&\nn \\
&{\cal A}^{<s,s} = P_s {\cal A} P_s = 
P_s ( a^s_0 \gamma_0 + a^s_1 1 + a^s_2 i\gamma_5 + a^s_3 \gamma_3 ). &\nn 
\eea
A consistent iterative solution of equations (\ref{sd1}) and (\ref{sd2}) yields the following 
equations for $s_0^s$ and $a_0^s$ (for details see~\cite{Kainulainen:2001cn})
\bea
&\hat C s_0^s = 0, \quad  \hat C a_0^s = 0,&\nn \\ 
&\hat K s_0^s = Coll., \quad  \hat K a_0^s = 0, &\nn 
\eea
with the constraint and kinetic operators 
\bea
&\hat C = k_0^2 - \vec k^2 - m^2 - s \frac{ m^2\theta^\prime}{\tilde k_0}, &
 \label{con} \\
&\hat K = \Diamond \{ \hat C, .\} = 2 k_3 \partial_{X_3} - (m^2)^\prime \partial_{k_3} 
-s\frac{ (m^2\theta^\prime)^\prime }{\tilde k_0} \partial_{k_3}.
  &\label{kin}
\eea

Analyzing the behaviour of these operators under CP conjugation,
we can identify the CP-even (index $^e$) and CP-odd  (index $^o$)
parts as follows,
\bea
&\hat C^e =k_0^2 - \vec k^2 - m^2 , \quad  \hat C^o 
   =- s \frac{ m^2\theta^\prime}{\tilde k_0}, &\nn \\ 
&\hat K^e =2 k_3 \partial_{X_3} - (m^2)^\prime \partial_{k_3}, 
  \quad  
\hat K^o =-s\frac{ (m^2\theta^\prime)^\prime }{\tilde k_0} \partial_{k_3}.
  &\nn 
\eea
The correctly normalized solutions for the spectral functions $a_0^s$ are
\bea
a_0^s &=& \pi i \, {\rm sign}(k_0) 
           (\partial_{k_0}\hat C) \, \delta (\hat C). \nn 
\eea
In this form we can immediately see, that the axial current will not contain
any CP violation if the wall velocity vanishes, since this would permit
the solution
\bea
s_0^s &=& f_F(k \cdot u=k_0)\, a_0^s \nn 
\eea
and the collision terms would vanish. This 
solution does not contain any CP violation:  
$(s_0^++s_0^-)$ is the zero component of the vector current and
odd under CP, while $(s_0^+-s_0^-)$ is the three component
of the axial current and even under CP.

The semi-classical picture of this process is clear. The wall profile 
gives a spin dependent and CP-violating dispersion relation to the fermions. 
But this effect 
can not lead to an axial current as long as the boundary conditions are invariant 
under a sign change of $k_3$. If the wall is moving, this symmetry is 
broken and this leads to an axial current. 
A good measure for the CP violation 
in the system in this context is the relative shift of the poles
of the dispersion relation (\ref{con})
\bea
\frac{\delta \omega}{\omega} = 
  s \frac{m^2\theta^\prime }{2 \tilde k_0 k_0^2} 
\,.
\label{delpol}
\eea

\section{Why the Standard Model (Na\"ively) Fails} \label{SMJarl}

In the Standard Model the Jarlskog 
determinant~\cite{Jarlskog:1985cw,Jarlskog:1988ii}
 is believed to be an upper bound on CP
violating effects. The basis for this relation is the following reasoning: 
Suppose the SM Lagrangian contains two nonhermitian mass matrices
for the quarks (due to the coupling to the Higgs field,
denoted by $\tilde m_u$ and $\tilde m_d$ )
 while the coupling of the lefthanded quarks to
the W bosons is still proportional to unity in flavour space.

Using four unitary flavour matrices for the left/right handed up/down quarks 
($U_u^L$, $U_u^R$, $U_d^L$, $U_d^R$)
the Lagrangian can be written in terms of the mass eigenstates 
($m_u = U_u^{L\dagger}\tilde m_u U_u^R $, 
 $m_d = U_d^{L\dagger}\tilde m_d U_d^R $). 
The unitary matrices for the 
right handed quarks have no physical significance, while the product of the 
left handed up/down matrices lead to the CKM matrix in the coupling term between
left handed quarks and W bosons ($ C = U_d^{L\dagger} \, U_u^L $). 

These Lagrangians are not in one-to-one 
correspondence: If we started with mass matrices, that needed the same left   
handed but different right handed transformation matrices, we would end up with the
same CKM matrix, and the same diagonal mass matrices. If we express now
our measurable quantities by the primary nondiagonal mass matrices, 
only combinations are allowed that do not include the right handed
transformation matrices after diagonalisation. 

In the SM the combinations of lowest dimension, that fulfill these 
requirements are the matrices 
$\tilde m_u \,\tilde m_u^\dagger$ and $\tilde m_d \,\tilde m_d^\dagger$, and it
turns out that the first CP sensitive contribution is the Jarlskog determinant 
\bea
&& \hskip -0.3in \Im \,\big( {\rm det}[\tilde m_u \,\tilde m_u^\dagger,
\tilde m_d \,\tilde m_d^\dagger]\big)
  \nn \\
&=& {\rm Tr}(C m_u^4 C^\dagger m_d^4 C m_u^2 C^\dagger m_d^2) \nn \\
&\approx& -2J \; m_t^4 m_b^4 m_c^2 m_s^2, \label{jarlsdet}
\eea
has dimension 12, and is suppressed by the 12th power of the W boson mass, 
or in a thermal system at least by the 12th power of the temperature.
On these grounds the first physical effect would be of order~\cite{Farrar:sp}
\bea
&\big(\frac{g_W^2}{2M_W^2}\big)^7 J \;
   m_t^6 m_b^4 m_c^2 m_s^2 \sim 10^{-22},& 
\label{shapbound}
\eea
where $J$ denotes a specific combination of the angles of the CKM
matrix~\cite{Jarlskog:1988ii,Hagiwara:fs}. For example, 
in the Kobayashi-Maskawa parametrization~\cite{KobayashiMaskawa:1973}
\begin{eqnarray}
V_{\rm CKM} =
  \Biggl(\begin{array}{ccc}
        c_1      \;&\; -s_1 c_3     \;&\; -s_1 s_3 \cr
        s_1c_2   \;&\; c_1 c_2 c_3 -s_2 s_3 {\rm e}^{i\delta} 
                     \;&\; c_1c_2s_3 + s_2 c_3 {\rm e}^{i\delta}  \cr
        s_1s_2   \;&\; c_1 s_2 c_3 + c_2 s_3 {\rm e}^{i\delta} 
                     \;&\; c_1s_2s_3 - c_2 c_3 {\rm e}^{i\delta}  \cr
        \end{array}
 \Biggr)
,
\nonumber
\end{eqnarray}
it is given by
\bea
J \!&=&\! s_1^2s_2s_3c_1c_2c_3\sin(\delta)
      = (3.0\pm0.3)\times 10^{-5}
\,,
\eea
where $s_i \equiv \sin(\vartheta_i)$ and $c_i\equiv \cos(\vartheta_i)\;$
($i=1,2,3$).

The calculation of the last section can as well be performed with 
several flavours. A numerical solution of the system shows, that 
the constraint equation (\ref{con}) contains a term 
\bea
&{\Im}\,\big({\rm Tr}[m_u \partial \, m_u^\dagger]\big)&
 \label{gen8}
\eea
as a generalization of (\ref{delpol}).
This term is only of dimension 3 and provides the possibility to circumvent the
upper bound (\ref{shapbound}) if one includes 
contributions that can produce these
terms, e.g. corrections due to the thermal self-energies. 
This inclusion is needed since the derivatives of the mass matrices 
are proportional to the mass matrices themselves.
Therefore the generalization (\ref{gen8}) can have no contribution on tree 
level. Even more stringent is the 
prejudice~\footnote{
A notable exception is Ref.~\cite{Farrar:sp},
in which the starting point is not the Kadanoff-Baym equations~(\ref{KB}),
but rather an effective action which includes one loop self energy 
corrections. A reflection calculation off a thin (step) wall 
performed in the mass eigenbasis then results in
CP violating quantities, which can be much larger than the Jarlskog invariant,
but which are in general not rephasing invariant. 
An important difference between the two treatments is that we have obtained
our result by working in the flavour eigenbasis; rotating the 
Kadanoff-Baym equations~(\ref{sd2}) into the mass eigenbasis would namely
lead to nontrivial derivative corrections~\cite{Kainulainen:2001cn}
which are not present in the effective action approach. 
}
that, in an expansion 
of the self-energy in masses, the 
most important 
contribution will be of the form (\ref{jarlsdet})
 such that the bound (\ref{shapbound}) 
still seems to hold. However, this argument is based on the assumption, that the 
convergence of an expansion in the mass parameters is fast, and this 
turns out not to be the case.    
  
\section{Self-energies in the Standard Model} \label{SMSelf}

The hermitian part of the thermal self-energy of the quarks in the Standard 
Model reads~\cite{Quimbay:1995jn}
\begin{eqnarray}
 \Sigma_R  &=& {k}\!\!\!/\, ( K_L \, P_L + K_R \, P_R) \nonumber \\
  && + {u}\!\!\!/\, ( U_L \, P_L + U_R \,P_R) \nonumber \\
  && + M \, P_L + M^\dagger \, P_R, 
\end{eqnarray}
with $K_L, K_R, U_L, U_R$ hermitian $3\times 3$ matrices, $M$ an arbitrary
$3\times 3$ matrix, all depending on $X_3$, the external energy 
$\omega=u \cdot k$,
and the external momentum $\kappa=\sqrt{\omega^2 - k^2}$ 
in the restframe of the plasma, 
$P_L, P_R$ the left/right-handedness 
projection operators and $u^\mu$ again the plasma vector.

In general all these coefficients can contain CP-violating contributions,
but we will focus on the mass term, since it leads to the generalisation 
of terms of the form (\ref{delpol}). The mass part of the thermal self-energy 
of the down quarks in the mass eigenbasis has the form
\bea
 M_d &=& h_1 m_d + \alpha_w C \frac{m_u^2}{m_W^2} h_2\, C^\dagger m_d \nn \\
  &&+ \quad \alpha_w^2 \int C \frac{m_u^2}{m_W^2} h_3\, C^\dagger 
    \frac{m_d^2}{m_W^2} h_4\, C \frac{m_u^2}{m_W^2}h_3\, C^\dagger m_d \nn \\
  &&+ \quad O(\alpha_w^3), \nn
\eea
where $h_1$ and $h_4$ depend only on $m_d^2$, while 
$h_2$ and $h_3$ depend on $m_u^2$. The integral is performed over 
the energies and momenta of the particles in the loop. 
The terms including the CKM matrices 
result only from the loops of the charged Higgs bosons and 
are displayed in fig.~(\ref{loops}).
\begin{figure}[htbp]
\begin{center}
\epsfig{file=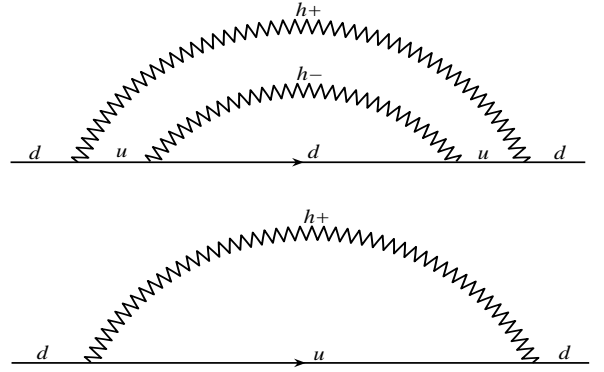, height=2.2in,width=3.3in}
\end{center}
\vskip -0.25in
\lbfig{loops}
\caption[fig5]{%
\small
Leading contributions to the nondiagonal term in the self-energy at one 
and at two loop level. 
}
\end{figure}
Since the derivatives of
the mass matrices are proportional to the mass matrices themselves, 
in the combination $\Im \,( {\rm Tr}[M_d \partial M_d^\dagger])$ only 
the derivatives of the $h$ functions will contribute. 
Furthermore the first CP sensitive term has to 
include at least four CKM matrices, and using the 
relation~\cite{Jarlskog:1988ii}
\bea
&&\hskip -0.15in {\rm Tr}(C X_1 C^\dagger X_2 C X_3 C^\dagger X_4) 
\nn 
\\
&& \hskip 1.1in
 =\;
 -2J \sum_{ij} \epsilon_{ikl} X_1^k X_3^l \epsilon_{jmn} X_2^m X_4^n  
\nn
\eea
for diagonal matrices $X$ with the entries $X^i$ and J as in (\ref{shapbound})
we find the following contributions 
(in the following prime denotes differentiation with respect to the Higgs {\it vev})
\bea
&& \hskip -0.2in {\rm Tr} (M_d M_d^{\dagger\prime} ) \nn \\ 
&=& \frac{\alpha_w^3}{m_W^8} \int  {\rm Tr} (  C^\dagger m_d^2 \, h_4 \,  
     C \, m_u^2\, h_3 \, C^\dagger m_d^4\, C\,m_u^4\, h_3\, h_2^\prime ) \nn \\
&&+ \frac{\alpha_w^3}{m_W^8} \int {\rm Tr} ( C^\dagger m_d^2\, C\, m_u^2\, h_3^\prime\, 
     C^\dagger m_d^2\, h_4\, C\, m_u^4\, h_3\, h_2 )\nn \\
&&+  \frac{\alpha_w^3}{m_W^8}\int {\rm Tr} ( C^\dagger m_d^2\, C\, m_u^2\, h_3\, 
     C^\dagger m_d^2\, h_4^\prime\, C\, m_u^4\, h_3\, h_2) \nn \\
&&+  \frac{\alpha_w^3}{m_W^8}\int {\rm Tr} ( C^\dagger m_d^2\,  C\, m_u^2\, h_3\, 
     C^\dagger m_d^2\, h_4\, C\, m_u^4\, h_3^\prime\, h_2). \nn \\
&&+ \quad O(\alpha_w^4) \nn
\eea
We do not attempt to calculate the two loop contribution, but give qualitative 
arguments how the enhancement of CP-violating terms appearing 
in $h_2$ result from the one loop calculation.

The thermal propagators for the up quarks $S(p)$ and the Higgs bosons 
$D(p)$ in the Feynman gauge are given by(see~\cite{Quimbay:1995jn} for details of
the calculation)
\bea
S(p)&=&({p}\!\!\!/\,+m_u) \left( \frac{1}{p^2-m_u^2+i\epsilon} 
  + i\Gamma_F(p) \right), \nn \\
D(p)&=&  \frac{1}{p^2-m_h^2+i\epsilon} 
  - i\Gamma_B(p) , \nn 
\eea
with the thermal parts
\bea
\Gamma_F &=& 2\pi \delta(p^2-m_u^2)f_F(p \cdot u),  \nn \\
\Gamma_B &=& 2\pi \delta(p^2-m_h^2)f_B(p \cdot u),  \nn 
\eea
and the fermionic and bosonic distribution functions
\bea
f_F(p \cdot u) &=& \frac{1}{\exp{(p \cdot u / T)}+1}, \nn \\
f_B(p \cdot u) &=& \frac{1}{\exp{(p \cdot u / T)}-1}. \nn 
\eea

The $T=0$ contributions undergo renormalization and are absorbed 
into the bare parameters of the Lagrangian. The remaining 
hermitian terms lead to the following form of $h_2$
\bea
&& \hskip -0.2in h_2(\omega,\kappa) 
 = \int \frac{d^4p}{(2\pi)^3}
\left( \frac{\Gamma_B(p)}{(p + k)^2 - m_u^2} - 
   \frac{\Gamma_F(p + k)}{ p^2 - m_h^2} \right), \nn
\eea 
and after three elementary integrations to 
\bea
&& \hskip -0.2in h_2(\omega,\kappa)
 =  \frac{1}{\kappa} \int_0^\infty \frac{d|{\bf p}|}{2\pi}
 \Big( \frac{|{\bf p}|}{\epsilon_h} L_2(\epsilon_h,|{\bf p}|) f_B(\epsilon_h) \nn \\
&& \hskip 1.1in
 - \frac{|{\bf p}|}{\epsilon_u} L_1(\epsilon_u,|{\bf p}|) f_F(\epsilon_u)\Big)
 . \nn
\eea 
The functions $L_1$ and $L_2$ are defined by
\bea
\hskip -0.3in L_{1/2}(\epsilon,|{\bf p}|) 
 &=& \log \left( \frac
{\omega^2 - \kappa^2 \pm \Delta + 2 \epsilon \omega + 2 \kappa |{\bf p}|}
{\omega^2 - \kappa^2 \pm \Delta + 2 \epsilon \omega - 2 \kappa |{\bf p}|} \right) \nn \\
 &+& \log \left( \frac
{\omega^2 - \kappa^2 \pm \Delta - 2 \epsilon \omega + 2 \kappa |{\bf p}|}
{\omega^2 - \kappa^2 \pm \Delta - 2 \epsilon \omega - 2 \kappa |{\bf p}|}
\right), \nn
\eea
where
 $\omega$ and $\kappa$ are the energy and the momentum of the external particle
in the restframe of the plasma, 
  $\epsilon_h = \sqrt{{\bf p}^2 + m^2_h}$, $\epsilon_u = \sqrt{{\bf p}^2 + m^2_u}$ 
and $\Delta=m_u^2 - m_h^2$.

\medskip

For $| \omega^2 - k^2 - \Delta | > 2 | m_u |$ or
    $| \omega^2 - k^2 + \Delta | > 2 | m_h |$ both particles in the loop can be onshell, 
whilst otherwise not. This makes the function $h_2$ strongly dependent on the
two loop masses. In the fig. (\ref{fig1})-(\ref{fig3}) $h_2$ is plotted as 
a function of the Higgs 
{\it vev} for a set of fixed external energies $\omega$ and momenta $k$
and three different masses $m_u$ of the quark in the loop.
\begin{figure}[htbp]
\begin{center}
\epsfig{file=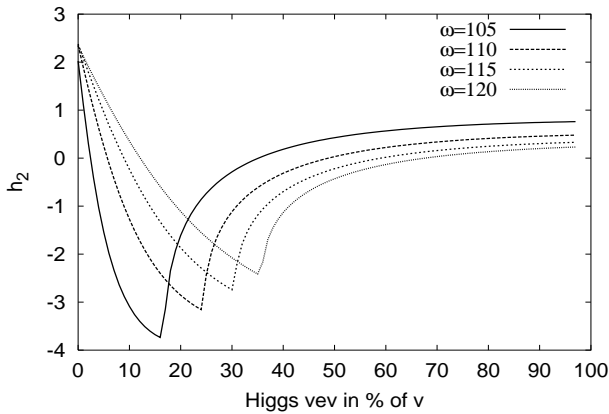, height=2.2in,width=3.3in}
\end{center}
\vskip -0.25in
\lbfig{fig1}
\caption[fig1]{%
\small
Dependence of $h_2$ on the Higgs {\it vev} $\langle\Phi\rangle$
in \% of its value $v=246~{\rm GeV}$ at $T=0$. 
The external energies and momenta are 
fixed at $\omega=105$ GeV to $\omega=120$ GeV, $k=100$ GeV, 
the mass of the quark in the loop is $m_u=100$ GeV.
}
\end{figure}
\begin{figure}[htbp]
\begin{center}
\epsfig{file=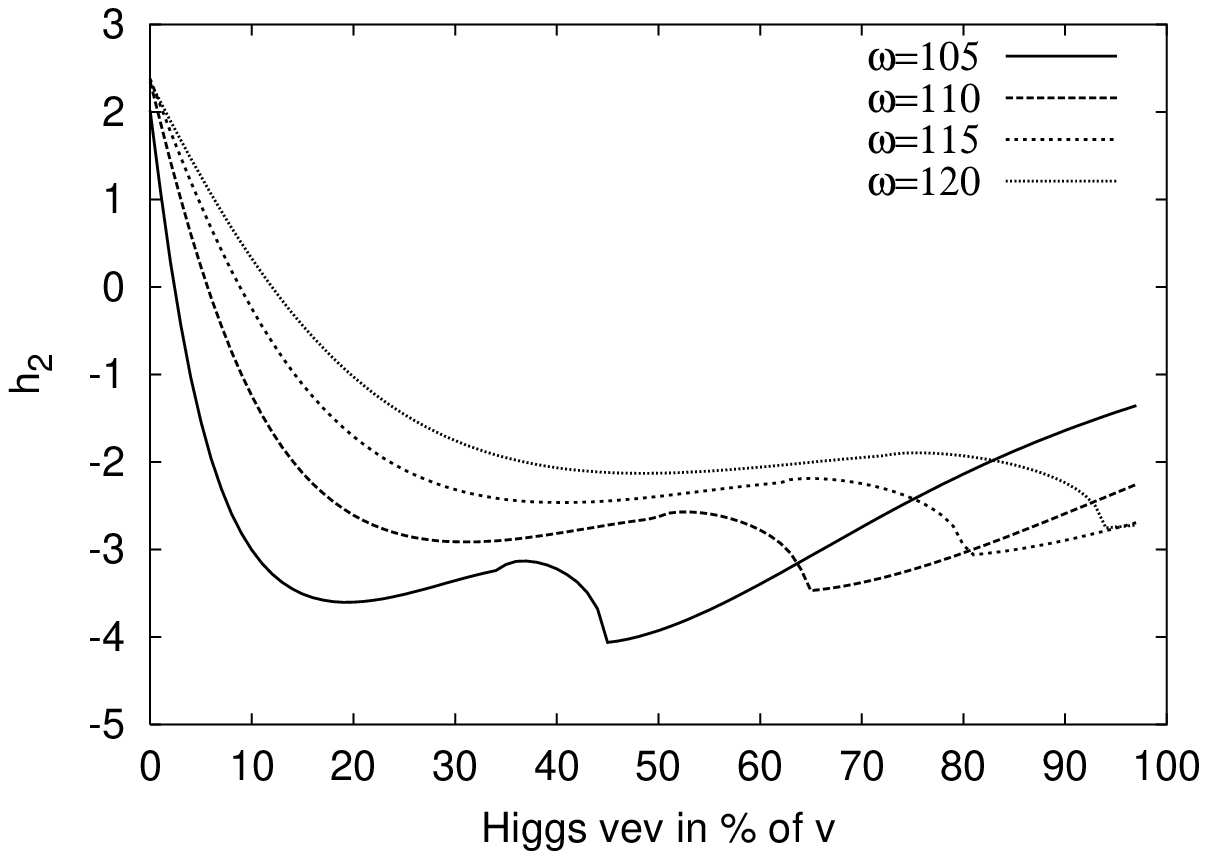, height=2.2in,width=3.3in}
\end{center}
\vskip -0.25in
\lbfig{fig2}
\caption[fig2]{%
\small
Same as fig.~(\ref{fig1}); mass of the quark in the loop is $m_u=10$ GeV.
}
\end{figure}
\begin{figure}[htbp]
\begin{center}
\epsfig{file=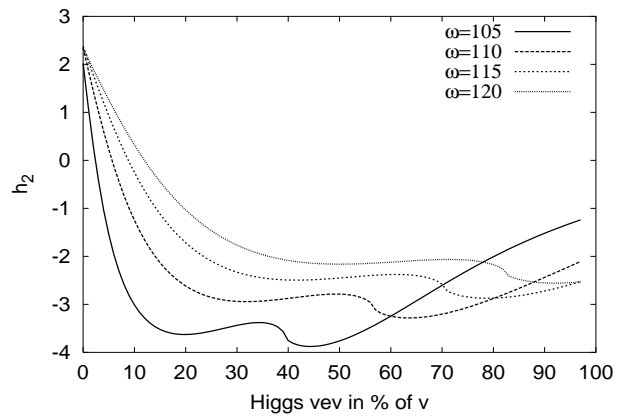, height=2.2in,width=3.3in}
\end{center}
\vskip -0.25in
\lbfig{fig3}
\caption[fig3]{%
\small
Same as fig.~(\ref{fig1}); mass of the quark in the loop is $m_u=1$ GeV.
}
\end{figure}
This strong dependence on the Higgs {\it vev} $\langle \Phi \rangle$ 
results in large 
derivatives of $h_2$ due to the wall profile. In fig.~(\ref{fig4})
$h_2^\prime$ is plotted versus the Higgs {\it vev} for different
internal quark masses.
\begin{figure}[htbp]
\begin{center}
\epsfig{file=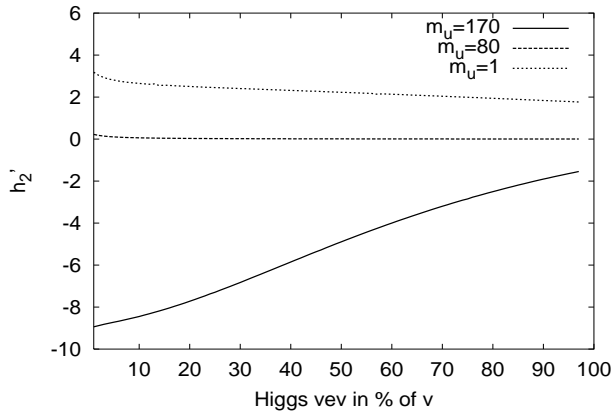, height=2.2in,width=3.3in}
\end{center}
\vskip -0.25in
\lbfig{fig4}
\caption[fig4]{%
\small
Dependence of $h_2^\prime$ on the Higgs {\it vev} with 
an on-shell external quark of mass $m_e=4$ GeV and an internal quark mass
in the range 1 GeV to 170 GeV. 
}
\end{figure}
Note that the first derivative of $h_2$ is in a broad range of parameter space
$\{\langle\Phi\rangle,m_u\}$ of order {\it unity} or 
even larger. Furthermore in the limit of vanishing external mass 
the sign of $h_2^\prime$ changes in the range,
where the internal quark mass agrees with the mass of the 
charged Higgs boson $m_h=m_W=80~{\rm GeV}$. 
The self-energy behaves non-perturbative in the 
sense that, when expanded in the mass of the internal quark, the main contributions 
come from higher powers of $m_u/m_W$. In fig.~(\ref{fig6}) the derivative 
$h_2^\prime$
is plotted versus the mass of the up quark in the loop. Here it is obvious that
the effect is based on a resonance in the loop and can not be increased 
arbitrarily by increasing the mass of the quark in the loop. 

\begin{figure}[htbp]
\begin{center}
\epsfig{file=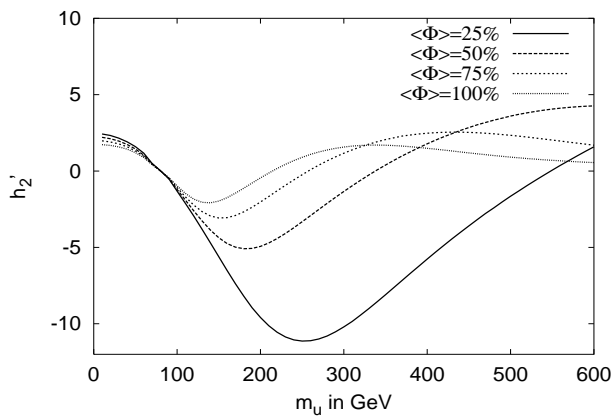, height=2.2in,width=3.3in}
\end{center}
\vskip -0.25in
\lbfig{fig6}
\caption[fig6]{%
\small
Dependence of $h_2^\prime$ on the mass of the quark in the loop with 
an on-shell external quark of mass $m_e=4$ GeV. The Higgs {\it vev} is
chosen in a range of 25\% to 100\% of its value in the broken phase at $T=0$. 
}
\end{figure}

It is reasonable to expect the functions $h_3$ and $h_4$ to be after integration 
effectively of order one as  well and not proportional to unity in flavour space.
This allows an estimate of the CP-violating pole dependence of the down quarks
$$ \frac{\delta\omega}{\omega} \sim J \;
m_t^4 m_s^2 m_b^2 m_c^2 \frac{ \alpha_w^3 h_2^\prime}
{m_W^8 l_w T^3} \sim 10^{-15},$$
(we have used the Standard Model value $l_w \, T \approx 20$ and that
most of the particles carry a momentum of the order of the temperature)
which is seven orders of magnitude larger than the
constraint~(\ref{shapbound}), but still much too small to account for the BAU. 
 
\section{Two Higgs Doublet Models}

We have seen, that in spite of the enhancement of the axial current, the 
CP-violating source due to the CKM matrix is too weak to account for the BAU. Thus we
will discuss in this section further possibilities to generate 
terms of the form~(\ref{delpol}) in extensions of the SM.

One attractive alternative is the extension to supersymmetric models.
Analytical~\cite{Carena:1996wj,Bodeker:1996pc,Espinosa:1996qw,Losada:1998at}
 and lattice~\cite{Laine:1998qk} studies 
show that the additional scalars in the theory of MSSM
may lead to a (two stage, see however \cite{Cline:1999wi})
first order phase transition in a part of the parameter space
fulfilling the sphaleron bound. 
However the occuring CP violation in the chargino 
and neutralino sector has to be maximal 
and the Higgs mass at the borderline to be seen experimentally
to explain the observed BAU~\cite{Cline:1997vk}.
Furthermore the current accuracy in the measurements of electric dipole
moments~\cite{Khalil:2002,Lebedev:2002,AbelKhalilLebedev:2001,
Pilaftsis:2002,Harris:jx,RomalisGriffithFortson:2000,LamoreauxGolub:1999}
only leaves a small window in parameter space of the MSSM 
and could rule out this model 
soon as a source of baryogenesis. An NMSSM type model~\cite{Huber:2000mg}
leaves more freedom at the expense of further parameters. 

Besides supersymmetric models, more general two Higgs doublet models 
are most appealing
in extending the SM to explain the BAU {\it via} electroweak baryogenesis.
In a certain region of parameter space according to Ref.~\cite{Cline:1996mg}, the
phase transition is of first order and the sphaleron bound 
is fulfilled.
This derivation assumed the two Higgs {\it vev}s to be proportional
to each other thus simplifying the calculation while the general case
so far has not been completely studied. 
Baryogenesis is not compatible with this assumption, since
if the quotient of the two Higgs {\it vev}s is constant,
it is not possible to generate an axial fermion current 
{\it via} Ref.~(\ref{delpol}). Therefore the character of the phase
transition has to be examined in every specific model seperately.   

Since these models are not subject to the stringent restrictions 
of supersymmetry, there are several possibilities to introduce
new sources of CP violation. 
One feasible approach is to avoid flavour changing neutral currents (FCNC) by
construction, what leads to the so called type I and type II models
(for a comprehensive discussion see Ref.~\cite{HiggsH}).
An additional source of CP violation in this context is the complex phase
between the two Higgs 
fields~\cite{McLerranShaposhnikovTurokVoloshin:1990,TurokZadrozny:1990,
DineHuetSingleton:1991,JoyceProkopecTurok:1994b,Bernreuther:2002uj}. 
Similarly to the supersymmetric case this model can, with a
reasonable choice of parameters, just marginally explain the
generated BAU to be consistent with primordial 
nucleosynthesis~\cite{Cline:1995dg}.

Another possibility, even if less attractive because of minor 
predictivity, is to admit FCNCs at the tree level, called 
two Higgs doublet models type III. Due to the large 
parameter space these models still resist to be ruled out
by experiments 
(for some implications on experimantal bounds see \cite{Atwood:1996vj})
even with quite natural choices for the new parameters and impressive
experimental lower bounds on FCNC processes.
The rich phenomenology can even account for deviations from
the SM as for example the difference between the measurement of
the $g-2$ muon factor and it's SM prediction~\cite{DiazRodolfo:2000yy}. 

The main difference to models without FCNC and particularly the SM is, 
that during the electroweak phase transition 
the derivatives of the mass matrices are not necessarily proportional to the 
mass matrices themselves. This gives the possibility to construct 
CP odd rephasing invariants of the form (\ref{gen8}) on the tree 
level and even with just two flavours.  

The Lagrangian considered for the the Yukawa couplings of the Higgs fields
to the quarks is of the form
\bea
\mathcal{L}_Y^{(III)}&=&\eta_{ij}^U \bar Q_{i,L} \tilde \phi_1 U_{j,R} 
+\eta_{ij}^D \bar Q_{i,L} \phi_1 D_{j,R}  \\
&+&\xi_{ij}^U \bar Q_{i,L} \tilde \phi_2 U_{j,R} 
+\xi_{ij}^D \bar Q_{i,L} \phi_2 D_{j,R} \nn
+ h.c.
\,,
\eea
where we used the standard notation: $Q_{i,L}$ denote the left-handed 
quark doublets, $U_{j,R}$ and $D_{j,R}$ ($i,j=1,2,3$) 
the up and down quark singlets, and $\phi_1$, $\phi_2$ 
are the two Higgs doublets. 
To fulfill the experimental bounds, it is sufficient to assume a  
hierarchy between the couplings $\eta^{U,D}$ and $\xi^{U,D}$.
In the basis where only the
Higgs field $\phi_1$ aquires a {\it vev} and after diagonalization of 
the fermion masses
the Yukawa couplings are parametrized \cite{Atwood:1996vj} as
\bea
\hat \eta^{U,D}_{ij} &=& \frac{m_i\delta_{ij}}{v}, \nn \\ 
\hat \xi^{U,D}_{ij} &=& \lambda_{ij} \frac{\sqrt{m_i m_j}}{v}, \nn \\ 
\eea
$|\lambda_{ij}| \lesssim 10^{-1}$ is needed to suppress $D^0 - \bar D^0$ and 
$B^0 - \bar B^0$ mixing sufficiently. 
Note that a change in the quotient of the two Higgs {\it vev}s in the mass 
eigenbasis 
of the Higgs fields leads to a change in the Yukawa couplings 
$\eta$ and $\xi$ in the above used basis with only one Higgs {\it vev}
and therefore to terms of the form~(\ref{gen8}).
The effect can be for example quite large in a two-stage 
phase transition, as it was seen in Ref.~\cite{Land:1992sm,Hammerschmitt:fn}.
Baryogenesis in these models occurs at the second phase transition, and it is 
efficient provided the first phase transition is sufficiently weak, 
such that the baryon number violating processes are not too suppressed 
in the weakly broken phase. 

The resulting pole shift will be of order 
\bea
\frac{\delta \omega}{\omega} &\approx& |\lambda|
   \frac{m^2}{4 k_0^3 l_w}, 
\eea
and for the top quark this is approximately 
\bea
\frac{\delta \omega}{\omega} &\sim&
   \frac{|\lambda|}{ T l_w},
\eea
with $T$ the temperature and $l_w$ the wall thickness. 
The high degree of arbitrariness in these models
opens this way a large window for electroweak baryogenesis.

\section{Conclusions}

We have shown that, due to a resonance in the quark self-energies, 
the CP-violating pole shift induced by the CKM matrix at high temperatures 
can be by about seven orders of magnitude larger than the CP-violating shift
na\"\i vely expected from the Jarlskog invariant. 
However, the effect is still too small to account for the BAU 
{\it via} baryogenesis from the Standard Model CP violation
at the electroweak phase transition.

Finally we point out that smallness of the CP violation 
in the Standard Model can be resolved within a certain class of
two Higgs doublet models.

\section*{Acknowledgements}

T.P. wishes to thank Jim Cline and Kari Rummukainen for discussions concerning 
the two Higgs doublet model. We thank Misha Shaposhnikov for a useful 
correspondence.

%
%

\end{document}